\documentstyle[twocolumn,aps,prl,floats,psfig]{revtex}
\begin{document}
\tightenlines
\twocolumn[\hsize\textwidth\columnwidth\hsize\csname@twocolumnfalse\endcsname
\title{Strong Tunneling and Coulomb Blockade in a Single-Electron Transistor}
\author{D. Chouvaev$^1$, L.S. Kuzmin$^1$, D.S. Golubev$^{1,3}$ and A.D.
        Zaikin$^{2,3}$}
\address{$^1$ Chalmers University of Technology, Fysikgr\"and 3,
         S-41296 Gothenburg, Sweden\\
         $^2$ Institut f\"ur Theoretische Festk\"orperphysik,
         Universit\"at Karlsruhe, 76128 Karlsruhe, Germany\\
         $^3$ I.E. Tamm Department of Theoretical Physics, P.N. Lebedev
         Physics Institute, Leninski pr. 53, 117924 Moscow, Russia}
\maketitle

\begin{abstract}
We have developed a detailed experimental study of a single-electron
transistor in a strong tunneling regime. Although
weakened by strong charge fluctuations, Coulomb effects were found
to persist in all samples including one with the effective
conductance 8 times higher than the quantum value (6.45 k$\Omega$)$^{-1}$.
A good agreement between our experimental data and theoretical
results for the strong tunneling limit is found.
A reliable operation of transistors with conductances 3-4 times larger
than the quantum value is demonstrated.

\end{abstract}
]


Mesoscopic tunnel junctions between metals represent a nontrivial
example of a macroscopic quantum system with discrete charge states
and dissipation \cite{AL,SZ}. Charging effects in such systems
can be conveniently studied in the so-called SET (single electron
tunneling) transistors. A typical SET transistor consists of a
central metallic island connected to the external leads via two
tunnel junctions with resistances $R_{L,R}$ and capacitances $C_{L,R}$
(see Fig. 1). In addition to the transport voltage $V$, the gate
voltage $V_g$ can also be applied to the metallic island via the gate
capacitance $C_g$.

\begin{figure}[h]
\centerline{\psfig{file=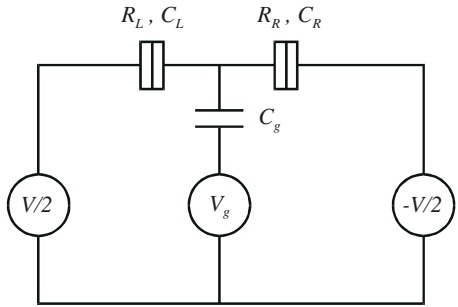,height=4cm}}
\caption{Schematics of a SET transistor.}
\label{fig1}
\end{figure}

Provided the junction resistances are large $R_{L,R} \gg
R_q=\pi \hbar /2e^2 \simeq 6.45$ k$\Omega$
tunneling effects are weak and can be treated perturbatively \cite{FN}.
For a quantitative measure of the tunneling strength we define the
parameter $\alpha_t=R_q/R_0$, where
$1/R_0=1/R_L+1/R_R$.
After each electron tunneling event
the charge of a central island changes exactly by $e$
and the energy difference between initial and final states of
a SET device is typically of order $E_C=e^2/2C$, where $C=C_L+C_R+C_g$.
As long as
tunneling is weak $\alpha_t \ll 1$ dissipative broadening of different
charge states $\Gamma$ is small (at low $T$ it is roughly
$\Gamma \sim \alpha_tE_C$) and these states are well resolved in energy.
This ensures nearly perfect quantization of the charge on
a central island in units
of $e$. As a result at sufficiently low $T \lesssim E_C$ Coulomb effects
dominate the behavior of a SET transistor leading to a number of
observable effects, such as Coulomb blockade of tunneling, modulation
of the current through a SET transistor by a gate voltage $V_g$,
Coulomb staircase on the $I-V$ curve etc. \cite{AL}.

The situation changes if the effective resistance $R_0$ becomes of order of
$R_q$ or smaller, i.e. $\alpha_t \gtrsim 1$. In this case dissipation is
large and the excited charge states of the system become broadened and overlap.
Do strong charge fluctuations lead to a complete smearing of Coulomb effects in
highly conducting mesoscopic tunnel junctions?

A positive answer on this question was suggested in Refs. \cite{Zw,BS}.
It was argued that -- similarly to Ohmic resistors
(see e.g. \cite{SZ}) -- charging effects should be {\it completely}
destroyed in junctions with $\alpha_t \gtrsim 1$. According to
\cite{Zw,BS} no Coulomb gap can exist for such values of $\alpha_t$
and a crossover to a purely Ohmic behavior was predicted.

Note that the above conclusion was obtained from studies of a certain
correlation function which turned out to be irrelevant for the problem
in question. In contrast, in Ref. \cite{PZ91} a nonperturbative analysis
of the junction ground state energy has been developed. This analysis
clearly demonstrates the existence of a nonvanishing Coulomb
gap $E_C^* \propto E_C \exp (-2\alpha_t)$ in the spectrum
of the system even for large $\alpha_t \gg 1$. Thus strong tunneling
{\it does not} destroy Coulomb effects, it only leads to effective
renormalization of the junction capacitance. As a result the temperature
interval relevant for charging effects shrinks, but they
still remain observable even at $T \gg E_C^*$.

This behavior is {\it qualitatively} different from that of an
Ohmic resistor. The physical reason for this difference
was also pointed out in \cite{PZ91}. It is due to different
symmetries of the allowed charge states: the symmetry is continuous
in the case of an Ohmic shunt, whereas only discrete $e$-periodic
charge states are allowed in the case of a normal tunnel junction
(see also Ref. \cite{SZ}). The latter symmetry remains the same for
any SET strength, and therefore at low $T$ Coulomb effects survive
and can be well observed even in highly conducting junctions.
The results \cite{PZ91} were supported by subsequent theoretical
studies (see e.g. \cite{GKS} for a discussion and further references).

In spite of all these theoretical developments, an experimental
investigation of this problem was
lacking. Recently Joyez et al. \cite{Esteve} carried out
an experiment aimed to study strong tunneling effects in SET
transistors with higher junction conductances. Deviations from
the standard ``orthodox'' theory \cite{AL} were clearly detected
in four different samples \cite{Esteve}. A very recent analysis \cite{KSS,GKS}
demonstrated that for three samples \cite{Esteve} with smaller $\alpha_t$
a remarkably good quantitative description of the data \cite{Esteve} is
achieved already within the perturbation theory in $\alpha_t$ if one retains
the ``cotunneling'' terms $\propto \alpha_t^2$. Hence,
the above samples \cite{Esteve} are still well in the
{\it perturbative} weak tunneling regime: the corresponding
values of $\alpha_t$ are large enough to observe deviations from the
``orthodox'' theory, but still too small for nonperturbative
effects to come into play (see below).

The main goal of the present paper is to develop a detailed
experimental study of Coulomb effects in mesoscopic tunnel junctions in
a {\it nonperturbative} strong tunneling regime  $\alpha_t \gtrsim 1$.
Beside its fundamental importance the problem is also of interest
in view of possible applications of SET transistors as electrometers.
The operating frequency range of such devices with $\alpha_t \ll 1$
is usually restricted to very low values. One possible way
to increase such frequencies is to decrease the junction resistance,
i.e. to fabricate SET transistors with $\alpha_t\gtrsim 1$.
But do such SET transistors exhibit charging effects?

In this paper we investigate if Coulomb effects
survive in the strong tunneling regime $\alpha_t \gtrsim 1$, in which
case discrete charge states are essentially smeared due to dissipation.
For this purpose we have carried out measurements of the current-voltage
characteristics of several SET transistors with
various values of the junction resistance $R_{L,R} \gtrsim R_q$
in the limit of a zero resistance of the external circuit.
The results are compared with the existing theory \cite{GKS,GZ}.

{\it Experiment}.
We have fabricated several SET transistors with different values of
the junction resistance.
The transistors were made using a standard electron-beam lithography
with two-layer resist and two-angle shadow evaporation of aluminum.
Five transistors with junction resistances in the range from 2 k$\Omega$
to 20 k$\Omega$ were studied. The corresponding values of $\alpha_t$ varied
between 1.5 and 8.3. The crossection
area of the tunnel junctions is estimated to be $\sim 0.01$ $\mu$m$^2$.

Measurements were done in a dilution refrigerator capable to hold temperature
from 20 mK to 1.2 K. Magnetic field of 2 Tesla was applied to keep aluminum in the normal state.
Thermocoax$^{\copyright}$ cables and a copper-powder
filter next to the mixing chamber of the refrigerator provided necessary
filtering against high frequency noise penetration.

We have measured the $I-V$ characteristics of the transistors at different
temperatures and gate voltages. At low voltages our measurements were
performed using a lock-in amplifier with 6 Hz reference signal frequency
in the constant voltage mode (a feedback was used to control the
amplitude of the excitation in order to keep the output signal at
the same level).

{\it Theory}.
At not very low temperatures (or voltages) the quantum dynamics of a tunnel
junction is well described by the quasiclassical Langevin equation \cite{AES}
\begin{equation}
C_j\frac{\hbar\ddot\varphi_j}{e}+\frac{1}{R_j}\frac{\hbar\dot\varphi_j}{e}=
J_j+\xi_{j1}\cos\varphi_j+\xi_{j2}\sin\varphi_j,
\label{Langevin}
\end{equation}
$j=L,R$; $J_j$ is the current flowing through the junction;
the phase variable $\varphi_j$ is related to the
voltage across the junction as $\hbar\dot\varphi_j/e=V_j$.
Discrete electron tunneling is responsible for the noise
terms in eqs. (\ref{Langevin}) described by the stochastic
Gaussian variables  $\xi_{jk}(t)$ with correlators
\begin{equation}
\langle\xi_{jk}(0)\xi_{j'k'}(t)\rangle =
\delta_{jj'}\delta_{kk'}\frac{\hbar}{R_j}
\int\frac{d\omega}{2\pi}e^{i\omega
t}\omega\coth\left(\frac{\hbar\omega}{2T}\right)
\end{equation}
The eqs. (\ref{Langevin}) are supplemented by the appropriate
current balance and Kirghoff equations and can be solved perturbatively
in the noise terms (see \cite{GKS,GZ} for details).
One arrives at the $I-V$ curve for a SET transistor \cite{GKS}:
\begin{eqnarray}
I(V)&=&\frac{V}{R_\Sigma}-I_0(V)-
\frac{V}{R_\Sigma}Ae^{-F}
\cos\left(\frac{2\pi Q_{\rm av}(V)}{e}\right);
\label{current}
\end{eqnarray}
where $Q_{\rm av}(V)=C_gV_g+\frac{R_LC_L-R_RC_R}{R_L+R_R}V$
is the average charge of the island and
$R_\Sigma=R_L+R_R$. The last two terms in
eq. (\ref{current}) describe deviations from the Ohmic
behavior due to charging effects. In the case $R_L=R_R$
for the $V_g$-independent term $I_0(V)$ we find \cite{GKS}
\begin{eqnarray}
I_0(V)=\frac{V}{8R_q}
\left[{\rm Re}\Psi (b)-{\rm Re}\Psi (a)\right]
-\frac{E_C}{\pi eR_0}{\rm Im}\Psi (b),
\label{I0}
\end{eqnarray}
$\Psi (x)$ is the digamma function, $a=1-ieV/4\pi T$
and $b=a+2\alpha_tE_C/\pi^2T$. The expression (\ref{I0}) holds for \cite{GKS}
\begin{equation}
\max\{eV,T\}\gg w_0= \left\{
\begin{array}{ll}
\frac{2\alpha_t}{\pi^2}{\rm e}^{-2\alpha_t+\gamma}E_C, & \alpha_t\gtrsim 1 \\
E_C, & \alpha_t\lesssim 1
\end{array}
\right.
\label{cond}
\end{equation}
$\gamma=0.577..$ is the Euler constant. The last term in eq. (3) describes
the modulation of the $I-V$ curve by the gate voltage. Provided the island
charge fluctuations are large the amplitude of the modulation is
exponentially suppressed  \cite{GKS,GZ}: $F(T,V) \gg 1$. The general
expression for the function $F(T,V)$ \cite{GKS} is quite complicated and
is not presented here. In the limit of small voltages and for
$T \ll \alpha_tE_C$ this expression becomes
simpler, and we get
\begin{equation}
F(T,0) \simeq (T/T_0)^2, \;\;\; T_0=\sqrt{12\alpha_t}E_C/\pi^2.
\label{supt}
\end{equation}
As the condition (\ref{cond}) should be simultaneously satisfied
eq. (\ref{supt}) makes sense only for $\alpha_t \gtrsim 1$.
The constant $A$ in eq. (\ref{current}) has the form \cite{GKS,GZ}
$A=f(\alpha_t)e^{-2\alpha_t}$. The prefactor $f(\alpha_t)$ was also
estimated in \cite{GKS,GZ} with a limited accuracy. More
accurate results for $f(\alpha_t )$ at low $T$ can be derived by means
of other techniques \cite{PZ91}.

{\it Results and Discussion}.
In what follows we will disregard a small asymmetry in the parameters of
$L$- and $R$-junctions and put $R_0=R_\Sigma/4$. The value of
$R_{\Sigma}$ was measured from the slope of the $I-V$ curve
at high voltages. The accuracy of these measurements was limited
by nonlinearities on the $I-V$ curves due to suppression of tunnel
barriers and heating and is estimated as $\sim$3\%.
The charging energy $E_C$ is usually determined from
the offset on the $I-V$ curve at large $V$. In the strong tunneling regime
$\alpha_t \gtrsim 1$ a clear offset can be reached only at very high
voltages where precise measurements are difficult due to other reasons.
Therefore, the above method gives only a rough estimate of $E_C$
with the accuracy within a factor 2. Alternatively one can try to determine
$E_C$ from the high temperature expansion of a zero-bias conductance:
$G(T)=(1/R_{\Sigma})(1-E_C/3T+..)$. For our samples this asymptotics
works well only at $T\gtrsim 10$ K and could not be reached in
the same cooling cycle. Therefore, the parameter $E_C$ was determined
from the best fit of the low-temperature ($\leq 1$K) and low-voltage
($\leq 700\mu$eV) parts of the $I-V$ curves averaged over the gate charge.
Fitting of $dI/dV$  for different temperatures allows to determine $E_C$
with a sufficiently high accuracy and avoid problems discussed above. Also
the precise value of $\alpha_t$ can be verified by this method.

The results are summarized in the following table:

\begin{table}[h]
\caption{Parameters of the samples}
\begin{tabular}{cddddd}
 Sample & $R_\Sigma$, k$\Omega$ & $\alpha_t$ & $E_c$, K & 10$w_0$,
 $\mu$eV & 10$w_0$, mK\\
 \hline
     I  &   17.4   &   1.48   &   2.25   &   50  &  600    \\
     II &   12     &   2.15   &   1.1    &   10  &  100  \\
    III &   10.4   &   2.48   &   1.04    &   5   &  60  \\
    IV  &   6.5    &   3.97   &   1.16   &   0.5 &  6   \\
    V   &   3.1    &   8.32   &   $\sim$ 0.3  & 5$\times 10^{-4}$ &  5$\times
10^{-3}$   \\
\end{tabular}
\label{table}
\end{table}

The last two columns show the lowest voltage and temperature $\sim 10w_0$
where the theory is still applicable.

\begin{figure}[h]
\centerline{\psfig{file=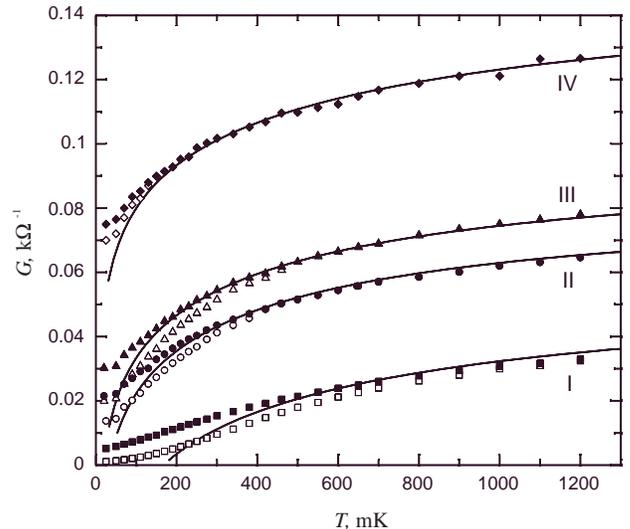,height=7cm}}
\caption{The averaged over $V_g$ (filled symbols) and
the minimum (open symbols) values for the linear conductance of the samples I-IV
together with a theoretical prediction for the averaged conductance (solid
curves).}
\label{fig2}
\end{figure}

The data for a temperature dependent zero bias conductance
averaged over the gate charge are given in
Fig.  2 for four samples (filled symbols) together with
the theoretical dependencies $G_{\rm av}^{\rm theor}=1/R_{\Sigma} -
I_0/V|_{V=0}$ (solid curves). The Coulomb blockade induced suppression of
the conductance at low $T$ is clearly seen even for a highly conducting
sample IV. We observe a good agreement
between theory and experiment except at the
lowest temperatures where the theoretical results become unreliable.
In this temperature interval the theoretical curves turn out to
be closer to the minimum conductance which is also shown by open symbols.
At low $T$ the system conductance shows a
tendency to saturation.  This is compatible with the corresponding conjecture
made in Ref.  \cite{GKS}.  On the other hand, heating effects as
an additional reason for this saturation cannot be excluded either.

\begin{figure}[h]
\centerline{\psfig{file=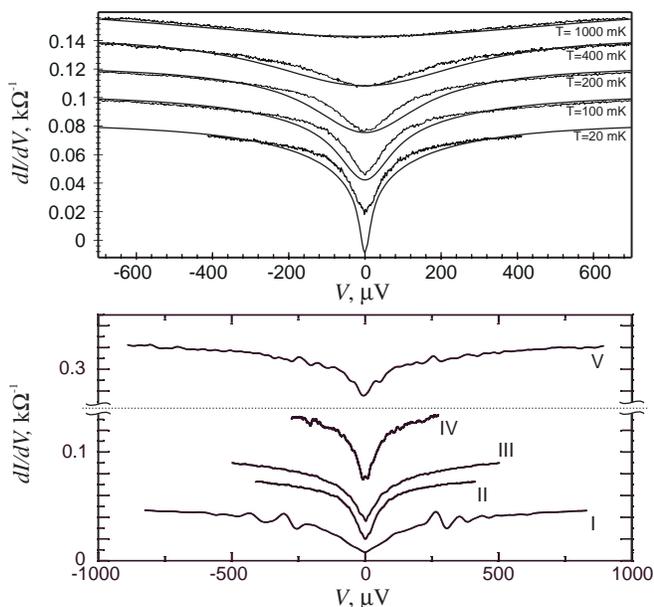,height=8cm}}
\caption{The $V_g$-averaged differential conductance.
Top panel: the data for the sample II (thinner curves) together
with theoretical results (thicker curves). For the sake of clarity the
curves at higher $T$ are shifted vertically with a step 0.02 k$\Omega^{-1}$.
Bottom panel: the data for all five samples at $T\simeq 20$ mK.}
\label{fig3}
\end{figure}

The averaged differential conductance $dI/dV$ is shown in the bottom panel
of the Fig.  3 for all five samples at the lowest temperature $T \approx 20$ mK.
It is remarkable that even for the sample V with $\alpha_t > 8$ a decrease of
the differential conductance at small $V$ due to charging effects is well
pronounced. The values $dI/dV$ measured for the sample II for different
temperatures are presented in the top panel together with a theoretical
prediction from eq. (\ref{I0}). A similarly good agreement between theory and
experiment was also found for the samples III, IV, and V.

\begin{figure}[h]
\centerline{\psfig{file=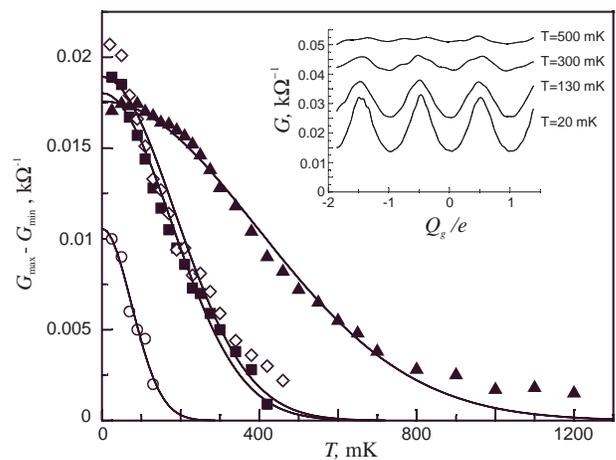,height=6cm}}
\caption{The gate modulation amplitude
$G_{\max}-G_{min}$ of a linear conductance as a function of $T$ for
the samples I-IV (triangles, squares, diamonds and circles respectively). Solid
lines provide the best fit with a formula $A\exp(-(T/T_0)^2)$. The inset: the
linear conductance of the sample II as a function of $Q_g=C_gV_g$ measured
at different $T$.} \label{fig4} \end{figure}

The gate modulation of the current was found to be of a simple cosine form
(\ref{current}) for all samples except for a relatively highly resistive
sample I at low $T$. The results for the gate modulated linear
conductance of the sample II are shown in Fig. 4a. The amplitude of
modulation increases with decreasing temperature in a qualitative
agreement with theory. At low $T$ the modulation effect is considerable
even for the sample IV with $\alpha_t \simeq 4$. At the lowest
temperatures this effect is visible also for the sample V. However
in this case the amplitude of modulation was only slightly above
the average noise level.

The data for the temperature dependent amplitude of conductance modulation
for the samples I-IV are presented in Fig. 4b. Solid curves correspond to
the best fit with a theoretical dependence $\propto \exp (-T^2/T_0^2)$
(cf. eqs. (\ref{current}, \ref{supt})). Note, that for all samples the
best fit value $T_0$ was found to be by a numerical factor $\sim 2\div 3$
smaller than the value (\ref{supt}). In other
words, the temperature suppression of the gate modulation is
{\it bigger} that it is predicted theoretically. We speculate
that this may be due to an additional effect of noise. Another possible
reason for such a discrepancy is an insufficient accuracy of the theoretical
calculations of the gate modulated conductance.

Our experimental results clearly
show that -- in accordance with
earlier theoretical predictions -- Coulomb blockade is {\it not}
destroyed even in the strong tunneling regime: clear signs of
Coulomb suppression of the transistor conductance were
observed for $\alpha_t$ as large as 8.3. For all $\alpha_t$ the
characteristic energy scale for charging effects is set by the
renormalized Coulomb gap $E_C^*$, however such effects remain
well pronounced even at $T \gg E_C^*$. The $I-V$ curves measured for
all five SET transistors are in a quantitative agreement with
the strong tunneling theory \cite{GKS,GZ}, except at very low
temperatures where this theory is not applicable. Along with the
overal suppression of the conductance its modulation by the gate
voltage was also observed at sufficiently low $T$. The modulation
effect increases with decreasing temperature in a qualitative
agreement with theoretical predictions.

Our work was aimed to study SET transistors in the nonperturbative
strong tunneling regime. In this sense our experiment is complementary
to that by Joyez et al. \cite{Esteve} where SET transistors with
smaller $\alpha_t$ were studied. Theoretically, a perturbative regime
characterized by an expansion parameter $g=\alpha_t/\pi^2 \ll 1$ (see
e.g. \cite{FN,GKS,KSS}) can be easily distinguished from a strong
tunneling one with $\exp (-2\alpha_t) \ll 1$ (cf. e.g. \cite{PZ91,GKS})
except within the interval $0.5 \lesssim \alpha_t \lesssim 3$ where both
inequalities are (roughly) satisfied. Combining our results with
those of Ref. \cite{Esteve} as well as with the corresponding theoretical
predictions \cite{GKS,KSS,GZ} we can draw a more definitive
conclusion about the validity range for both regimes: tunneling
can be treated perturbatively for $\alpha_t \lesssim 1$ while
the nonperturbative tunneling regime sets in for $\alpha_t \gtrsim 2$.
In the intermediate $1\lesssim \alpha_t \lesssim 2$ nonperturbative
results can be also applied at temperatures not much lower than $E_C$.

In summary, we have demonstrated that SET transistors operate well even
if the effective tunneling resistance $R_{0}$ is {\it several times smaller} 
than 6.5 k$\Omega$.

The work was supported by
the Deutsche Forschungsgemeischaft within SFB 195, by the INTAS-RFBR
Grant 95-1305, by the Swedish NFR and by SI. The samples were
fabricated at the Swedish Nanometer Laboratory.



\begin{references}
\bibitem{AL} D.V. Averin and K.K Likharev, in {\it Mesoscopic Phenomena in
             Solids}, ed. by B.L. Altshuler, P.A. Lee and R.A. Webb, eds., p.
             173 (Elsevier, Amsterdam, 1991).
\bibitem{SZ} G. Sch\"on and A.D. Zaikin, Phys. Rep. {\bf 198}, 237 (1990).
\bibitem{FN} In this regime nonperturbative effects become important
only in the vicinity of the Coulomb blockade threshold, see
K.A. Matveev, Zh. Eksp. Teor. Fiz. {\bf 99}, 1598 (1991) [Sov. Phys.
JETP {\bf 72}, 892 (1991)]; D.S. Golubev and A.D. Zaikin, Phys.
Rev. B {\bf 50}, 8736 (1994);  H. Schoeller and G. Sch\"on, Phys.
Rev. B {\bf 50}, 18436 (1994).
\bibitem{Zw} W. Zwerger and M. Scharpf, Z. Phys. {\bf 85}, 421 (1991).
\bibitem{BS} R. Brown and E. Simanek, Phys. Rev. B {\bf 45}, 6069 (1992).
\bibitem{PZ91} S.V. Panyukov and A.D. Zaikin, Phys. Rev. Lett. {\bf 67},
3168 (1991).
\bibitem{GKS} D.S. Golubev, J. K\"onig, H. Schoeller,
G. Sch\"on and A.D. Zaikin, Phys. Rev. B {\bf 56}, 15782 (1997).
\bibitem{Esteve} P. Joyez, V. Bouchiat, D. Est\`eve, C. Urbina
and M.H. Devoret, Phys. Rev. Lett. 79, 1349 (1997).
\bibitem{KSS} J. K\"onig, H. Schoeller, and
G. Sch\"on, Phys. Rev. Lett. {\bf 78}, 4482 (1997).
\bibitem{GZ} D.S. Golubev and A.D. Zaikin, Zh. Eksp. Teor. Fiz. Pis'ma Red.
             {\bf 63}, 953 (1996) [JETP Lett. {\bf 63}, 1007 (1996)].
\bibitem{AES} U. Eckern, G. Sch\"on and V. Ambegaokar, Phys. Rev B {\bf 30},
6419, (1984); D.S. Golubev and A.D. Zaikin, Phys. Rev. B {\bf 46}, 10903
(1992).

\end{references}
\end{document}